\documentclass[structabstract]{aa} % for 2 col 
\usepackage{longtable,booktabs}
\usepackage{latexsym}
\usepackage{amssymb}
\usepackage{lscape}
\usepackage{natbib}
\usepackage{subfigure}
\usepackage{xcolor}
\usepackage{multicol}
\usepackage{multirow}
\usepackage{amsmath}
\usepackage{booktabs}
\usepackage{caption}
\usepackage{pdflscape}
\usepackage{graphicx}
\usepackage{txfonts}
\def\lsim{\mathrel{\rlap{\lower 3pt \hbox{$\sim$}} \raise 2.0pt \hbox{$<$}}}
\def\gsim{\mathrel{\rlap{\lower 3pt \hbox{$\sim$}} \raise 2.0pt \hbox{$>$}}}

\title{A powerful (and likely young) radio-loud quasar at z=5.3}
\subtitle{}

\author{S. Belladitta\inst{1,2}                                
	\and A. Moretti\inst{1}  
	\and A. Caccianiga\inst{1}
	\and D. Dallacasa\inst{3,4} 
	\and C. Spingola\inst{4} 
	\and M. Pedani\inst{5}
	\and L. P. Cassar\`a\inst{6}
	\and S. Bisogni\inst{6}}

\institute{INAF $-$ Osservatorio Astronomico di Brera, via Brera, 28, 20121 Milano, Italy\\
	\email {silvia.belladitta@inaf.it}
	\and
	DiSAT, Universit\`a degli Studi dell'Insubria, Via Valleggio 11, 22100 Como, Italy
	\and
		Dipartimento di Fisica e Astronomia, Universit\`a degli Studi di Bologna, Via Gobetti 93/2, I$-$40129 Bologna, Italy 
	\and
		INAF $-$ Istituto di Radioastronomia, Via Gobetti 101, I$-$40129, Bologna, Italy
	\and
	INAF $-$ Fundaci\'on Galileo Galilei, Rambla Jos\'e Ana Fernandez P\'erez 7, 38712 Bre\~{n}a Baja, TF, Spain
	\and
	INAF $-$ Istituto di Astrofisica Spaziale e Fisica Cosmica (IASF), Via A. Corti 12, 20133 Milano, Italy
}

\begin{document}
	
	\date{Received; accepted}

	\abstract
	{We present the discovery of PSO~J191.05696$+$86.43172 (hereafter PSO~J191$+$86), a new powerful radio-loud quasar (QSO) in the early Universe (z = 5.32). 
	We discovered it by cross-matching the NRAO VLA Sky Survey (NVSS) radio catalog at 1.4~GHz with the first data release of the Panoramic Survey Telescope and Rapid Response System (Pan-STARRS PS1) in the optical.  
	With a NVSS flux density of 74.2~mJy, PSO~J191$+$86 is one of the brightest radio QSO discovered at z$\sim$5. 
	The intensity of its radio emission is also confirmed by the very high value of radio loudness (R>300).
	The observed radio spectrum of PSO~J191$+$86 shows a possible turnover around $\sim$1 GHz (i.e., $\sim$6 GHz in the rest frame), making it a Gigahertz-Peaked Spectrum (GPS) source. However, variability could affect the real shape of the radio spectrum, since the data in hand have been taken $\sim$25 years apart. 
	By assuming a peak of the observed radio spectrum between 1 and 2 GHz (i.e. $\sim$ 6 and 13 GHz in the rest-frame) we found a linear size of the source of $\sim$10-30 pc and a corresponding kinetic age of 150-460 yr. This would make PSO~J191$+$86 a newly born radio source. 
	However, the large X--ray luminosity (5.3$\times$10$^{45}$ erg s$^{-1}$), the flat X--ray photon index ($\Gamma_X$=1.32) and the optical-X--ray spectral index ($\tilde{\alpha_{ox}}$=1.329) are typical of blazars.
	This could indicate that the non-thermal emission of PSO~J191$+$86 is Doppler boosted. 
	Further radio observations (both on arcsec and parsec scales) are necessary to better investigate the nature of this powerful radio QSO.}  

	
	\maketitle
	%
	%________________________________________________________________

\section{Introduction}
\label{intro}
Radio-loud (RL) active galactic nuclei (AGN) are those supermassive black holes (SMBHs) that are able to expell part of the accreting matter into two relativistic bipolar jets (e.g., see \citealt{Blandford2019} for a recent review). 
Therefore they are usually called \emph{jetted} AGN \citep[e.g.,][]{Padovani2017} and they represent up to 10\% of the total AGN population \citep[e.g.,][]{Liu2021}. 
Understanding the mechanisms responsible for the launch and the emission of these jets is of crucial importance to study their role in SMBHs accretion and evolution \citep[e.g.,][]{Volonteri2015}, as well as to investigate their feedback on the intergalactic medium \citep[e.g.,][]{Fabian2012}.\\ 
High-redshift RL quasars (QSOs) are critical tools for studying the early evolutionary stage of the first jetted SMBHs, their feedback on the host galaxy and the environment, and their contribution to the re-ionization epoch \citep[e.g.,][]{Blandford2019}.\\
Among the more than 500 QSOs discovered to date at z$>$5, $\sim$30 are classified as RL\footnote{Usually an AGN is considered to be radio loud when it has a radio loudness R$>$10, with R defined as the ratio of the 5~GHz and 4400$\mbox{\AA}$ rest-frame flux densities: R = S$_{5 \rm GHz}$ /S$_{4400\AA}$ \citep{Kellermann1989}.} (e.g., \citealt{Romani2004,McGreer2006,Zeimann2011,Sbarrato2012,Banados2018}). 
Therefore, a systematic search and study of these objects is necessary to better constrain the properties of jetted SMBHs in the first Gyrs since the Big Bang.\\
% --- --- --- 
In an effort to enlarge the current sample of high-z RL QSO, we are conducting a project that combines optical, infrared, and radio datasets to identify distant radio sources all over the sky (\citealt{Caccianiga2019,Belladitta2019,Belladitta2020}, Ighina et al. in prep.). 
In this paper we present the discovery and the first observations of PSO~J191.05696$+$86.43172 (hereafter PSO~J191$+$86), a powerful jetted QSOs at z=5.32, which has been selected from the cross-correlation of the NRAO VLA Sky Survey (NVSS, \citealt{Condon1998}) in the radio, the Panoramic Survey Telescope and Rapid Response System (Pan-STARRS PS1, \citealt{Chambers2016}) in the optical and the AllWISE Source Catalog (\citealt{Wright2010,Mainzer2011}) in the mid-infrared (MIR).\\ 
The paper is organized as follows: in Sect. \ref{selection} the
selection method is outlined; in Sect. \ref{newobs} new optical and near-infrared (NIR) spectroscopic observations of PSO~J191$+$86 are presented;
in Sect. \ref{radioarchive} and Sect. \ref{swift} archival radio data and a new X-ray follow-up observation are reported, respectively; 
the results on the multi-wavelength properties of the source are described in Sect. \ref{optirprop} and \ref{results}; 
finally Sect. \ref{conc} reports a brief discussion and conclusions.\\
The magnitudes used in this work are all in the AB system, unless otherwise specified.
We use a flat $\Lambda$ cold dark matter ($\Lambda$CDM) cosmology
with H$_0$ = 70~km~s$^{-1}$~Mpc$^{-1}$, $\Omega_m$ = 0.3 and $\Omega_{\Lambda}$ = 0.7. 
Spectral indices are given assuming S$_{\nu} \propto \nu^{-\alpha}$ and all errors are reported at 1$\sigma$, unless otherwise specified.
Throughout the paper the flux densities are the observed ones, while luminosities are given in the source's rest-frame, unless otherwise specified.

\section{Source selection}
\label{selection}
\begin{figure*}[!h]
	\centering
	\includegraphics[width=19.0cm]{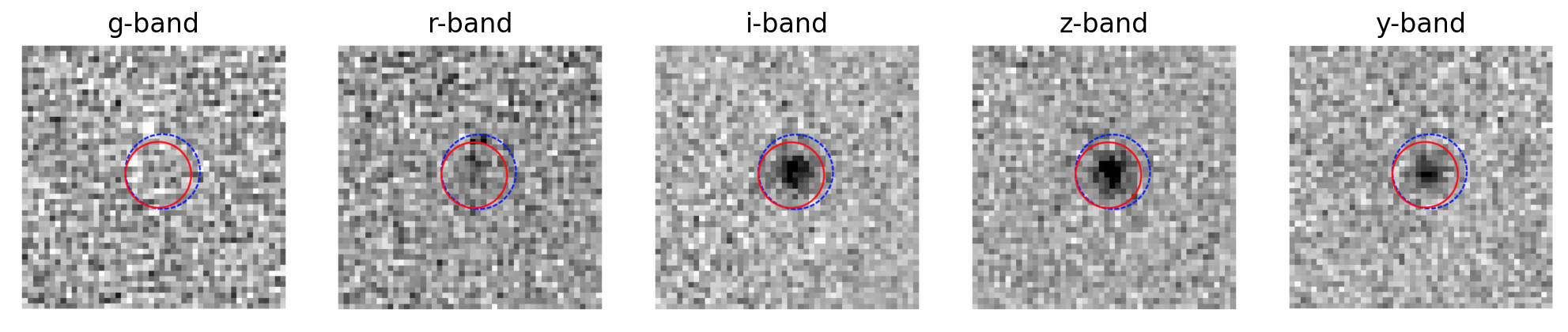}
	\vskip -0.2 true cm
	\caption{\small 0.2$'$$\times$0.2$'$ Pan-STARRS PS1 $g,r,i,z,y$ cutout images of PSO~J191$+$86. Its optical position is marked with a red circle of 1.5$''$ of diameter. The radio NVSS position is marked with a blue dashed circle, as large as the radio positional error reported in the catalog (1.7$''$). All images are oriented with north up and east to the left.}
	\label{psojphotimag}
\end{figure*}
From the entire NVSS catalog, we selected bright (S$_{1.4 \rm GHz}\geq$ 30 mJy) and compact objects to provide an accurate (<2$''$) radio position.
With this criteria we found $\sim$45000 sources. 
We then cross-matched them with the Pan-STARRS PS1 catalog, using a maximum separation equal to 2$''$. 
This impact parameter guarantees to recover more than 90\% of the real optical counterparts \citep{Condon1998}.
We selected only relatively bright optical sources (i$_{PS1}$ < 21.5) outside the Galactic plane (|b|$\geq$20$^{\circ}$), to minimize contamination from stars, and at Dec$>$-25$^{\circ}$, to exclude optical objects at the declination limit of Pan-STARRS survey.
Then all the sources that satisfied the following photometric criteria have been selected: \\
i) no detection in g$_{PS1}$-band; \\ 
ii) drop-out:  r$_{PS1}$-i$_{PS1}$ $\geq$ 1.2; \\
iii) blue continuum: i$_{PS1}$-z$_{PS1}$ $\leq$ 0.5; \\
iv) point-like sources: i$_{PS1}$-i$_{Kron}$ $<$ 0.05; \\
v) no detection in WISE (W2) or i$_{PS1}$ - W2(Vega) < 5.0. \\
This last constraint has been placed to minimize the contamination by dust reddened AGN at z = 1-2 (e.g., \citealt{Carnall2015,Caccianiga2019}).\\
Fourteen candidates remained after the application of these filters; four of them are already known radio QSOs in the literature. 
PSO~J191$+$86 stands out among the other 10 targets for its very high value of the drop-out ($r-i > 1.7$), its photometric redshift (z>5) and its very high radio flux density (S$_{1.4\rm GHz} >$ 70 mJy).
Therefore we considered only this source for a spectroscopic follow-up. \\
In Table \ref{psojmag} the Pan-STARRS PS1 and WISE magnitudes of PSO~J191$+$86 (corrected for Galactic extinction, using the extinction law provided by \citet{Fitzpatrick1999}, with R${_V}$ = 3.1) are reported.
Figure~\ref{psojphotimag} shows the optical images of the source in the different $g,r,i,z,y$ PS1 bands. 
The optical coordinates of PSO~J191$+$86 are: R.A. = 191.05696 deg (12$^h$44$^m$13.902$^s$), Dec = $+$86.43172 deg ($+$86$^d$25$^{'}$54.07$^{''}$). 
% --- --- 
\begin{small}
	\begin{table}[!h]
		\caption{\small Optical and MIR magnitudes of PSO~J191$+$86.}
		\label{psojmag}
		\centering
		\begin{tabular}{cccc}
			\hline\hline
					filter & central $\lambda$  & mag  & Ref. survey  \\  
		       & ($\mu$m)           & (AB) &              \\
		 (1)   & (2) & (3) & (4) \\
		\hline
		$r_{PS1}$ & 0.680 & 21.34 $\pm$ 0.15 & PS1 \\	 %21.92
		$i_{PS1}$ & 0.745 & 19.78 $\pm$ 0.04 & PS1 \\	%20.21
		$z_{PS1}$ & 0.870 & 19.49 $\pm$ 0.03 & PS1 \\ %19.83
		$y_{PS1}$ & 0.978 & 19.22 $\pm$ 0.06 & PS1 \\ %19.51
		$W1$ & 3.4 & 18.502 $\pm$ 0.035 & WISE \\ %15.819 
		$W2$ & 4.6  & 18.386  $\pm$ 0.051 & WISE \\ %15.067
			\hline
		\end{tabular}
		\tablefoot{Col (1): optical / MIR filters; Col (2): filter central wavelength in $\mu$m; Col (3): observed AB de-reddened magnitude; Col (4): reference catalog. The offset between the Pan-STARRS PS1 optical and WISE MIR positions is 0.85$''$. The relations to convert from Vega to AB systems are: $W1_{AB}$ = $W1$+2.683; $W2_{AB}$ = $W2$+3.319 \citep{Cutri2012}.}
	\end{table}
\end{small}

\section{Optical, NIR observations and data reduction}
\label{newobs}
\begin{figure*}[!h]
	\centering
	\includegraphics[width=15.0cm, height=7.5cm]{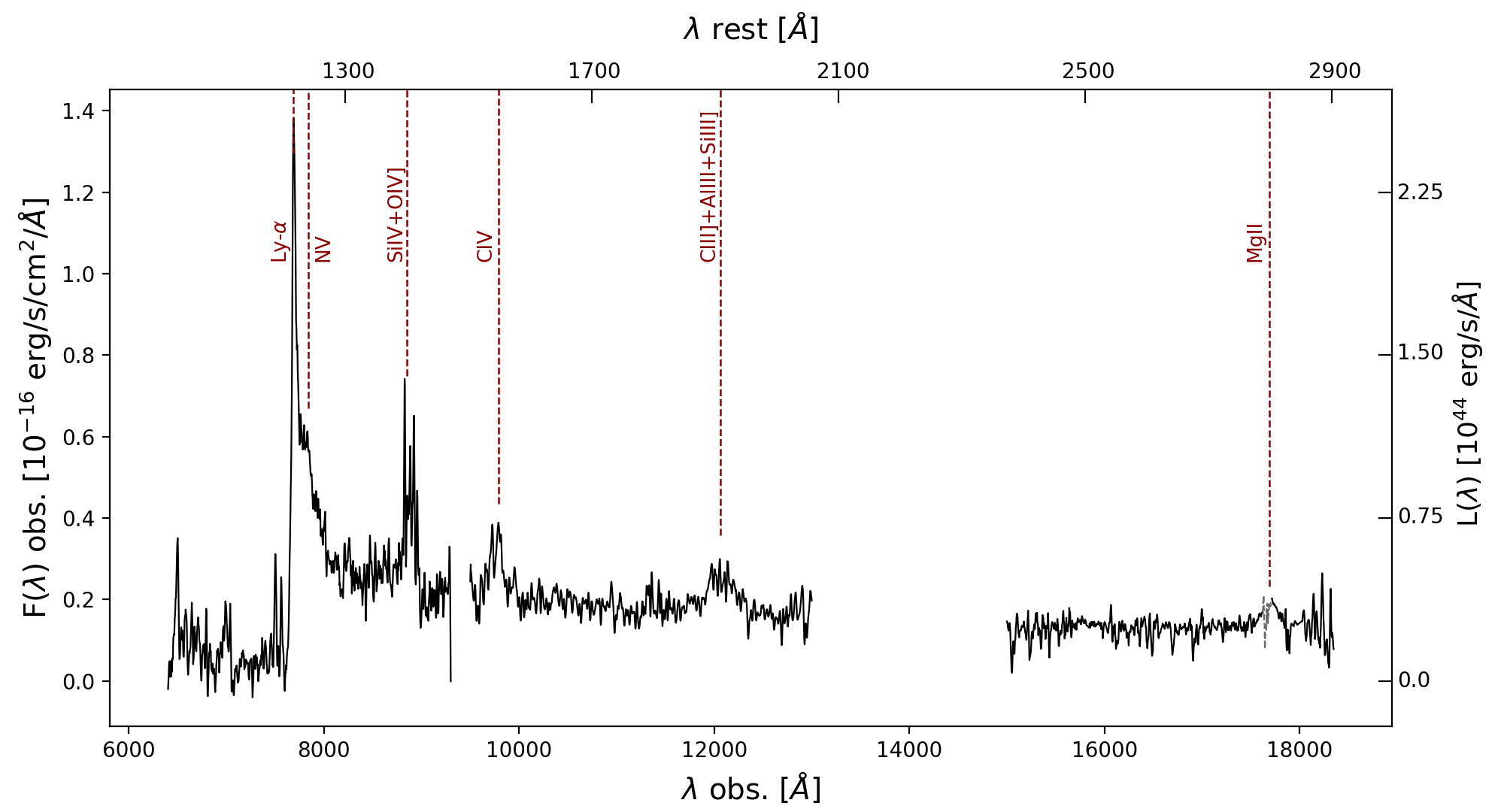}
	\caption{\small Optical and NIR de-reddened observed spectra of PSO~J191$+$86 taken at TNG and LBT respectively. The most important visible emission lines are marked. In grey we mask the part of the Mg$\rm II$ line that suffers from a strong absorption feature, in which the peak of the Mg$\rm II$ line falls.
	The top x-axis reports the rest-frame wavelengths while the right axis shows the monochromatic luminosity.}
	\label{psojspec}
\end{figure*}
\begin{small}
	\begin{table}[!h]
		\caption{\small Summary of the Optical and Near-infrared Follow-up Spectroscopic Observations of PSO~J191$+$86.}
		\label{sumobs}
		\centering
		\begin{tabular}{cccc}
		\hline\hline
		Date & Telescope/Instrument  & Exposure Time & S/N \\
		(1) & (2) & (3)  & (4) \\
		\hline
		2018 June 21     & TNG/DOLORES  & 0.5h  &  $\sim$7  \\ 
		
		2019 May 3     & LBT/LUCI1  &  1.8h &   $\sim$7.5    \\ 
		
		2019 May 3       & LBT/LUCI2  & 1.8h &  $\sim$7.5   \\
		\hline
		\end{tabular}
		\tablefoot{\small Col(1): Date of the observation; Col(2): Telescope and instrument used; Col(3): Totale exposure time; Col(4): Signal-to-noise ratio of the continuum.}
	\end{table}
\end{small}
\subsection{TNG/DOLORES observation}
We performed a dedicated spectroscopic follow-up of PSO~J191$+$86 with DOLORES (Device Optimized for the LOw RESolution) installed at the Telescopio Nazionale Galileo (TNG).
We confirmed PSO~J191$+$86 as a high-z RL QSO the night of the 21 June 2018, with a single 30~minutes observation with the LR-R grism and a long-slit of 1$''$ width. The mean air mass during the observation was 1.7. Table \ref{sumobs} reports the details of this observation. \\
An exposure of a Ar+Ne+Hg+Kr lamp was done to ensure the wavelength calibration and the flux calibration was obtained by observing the G191-B2B (R.A.=05$^h$05$^m$30.62$^s$, Dec=$+$52$^d$49$^{'}$54.0${''}$) spectro-photometric standard star of the catalogs of \citet{Oke1990}. \\
The data reduction was performed using standard Image Reduction and Analysis Facility (IRAF) procedures \citep{Tody1993}.
The DOLORES discovery spectrum is shown in Fig.~\ref{psojspec}; it has been corrected for Galactic extinction as explained in Sect.~\ref{selection}.

\subsection{LBT/LUCI follow-up}
\label{psolbtfollow}
We performed a spectroscopic follow-up with LBT utility camera in the infrared (LUCI, \citealt{Seifert2003}) at the Large Binocular Telescope (LBT) in order to extend the wavelength range in the NIR band and detect the C$\rm IV\lambda1549$ (hereafter C$\rm IV$) and the Mg$\rm II\lambda2798$ (hereafter Mg$\rm II$) broad emission lines (BELs), which can be used to estimate the central black hole mass of PSO~J191$+$86 thanks to the so-called single epoch (SE) or virial method (e.g., \citealt{Vestergaard2006}, hereafter VP06, see Sect. \ref{bhmass}).
The observation was carried out on the night of 2019 May 3 (P.I. Moretti A., program ID: LBT2018AC123500-1) and consisted of 12 exposures of 270s each in nodding mode in the sequence ABBA, with a total integration time of 1.8 hours.
The medium seeing along the night was 0.9$''$ and the mean air mass was 1.7.
The details of this observation are reported in Table \ref{sumobs}.
We decided to use the grism~200 with the zJ and HK filters on LUCI1 and LUCI2 respectively in order to cover all the spectral range and detecting the C$\rm IV$ (LUCI1) and the Mg$\rm II$ (LUCI2) simultaneously. 
The G200-zJ configuration allowed us to observe the wavelength range from 0.9 to 1.2 $\mu$m, where the C$\rm IV$ line is expected ($\lambda_{obs}$ = 9805$\mbox{\AA}$ at z$=$5.3); instead the G200-HK configuration covers the range from 1.5 to 2.4 $\mu$m in which the Mg$\rm II$ line falls (expected $\lambda_{obs}$ = 17700$\mbox{\AA}$). 
The data reduction was performed at the Italian LBT Spectroscopic Reduction Center, with the software developed for LBT spectroscopic data reduction \citep{Gargiulo2022}.
Each spectral image was independently dark subtracted and flat-field corrected. 
Sky subtraction was done on 2D extracted, wavelength calibrated spectra. Wavelength calibration was obtained by using several sky lines, reaching a rms of 0.25$\mbox{\AA}$ on LUCI1 and 0.5$\mbox{\AA}$ on LUCI2.\\
In Fig.~\ref{psojspec} we report the LBT/LUCI1 and the LBT/LUCI2 spectra in which the C$\rm IV$  and the Mg$\rm II$ BELs have been detected, together with the C$\rm III]\lambda$1909+Al$\rm III\lambda$1860+Si$\rm III]\lambda$1892 complex. 
Also in this case, the two spectra have been corrected for Galactic extinction as the TNG/DOLORES one. 
We note that the Mg$\rm II$ line suffers from the presence of an absorption feature (a residual of background subtraction, see grey line in Fig. \ref{psojspec} and the zoom-in in Fig. \ref{fitlines}). 
When masking the absorption feature, the wide range of absorption results in a gap close to the line center.

\section{Archival continuum radio data}
\label{radioarchive}
\begin{figure}[!h]
	\centering
\includegraphics[width=7.0cm]{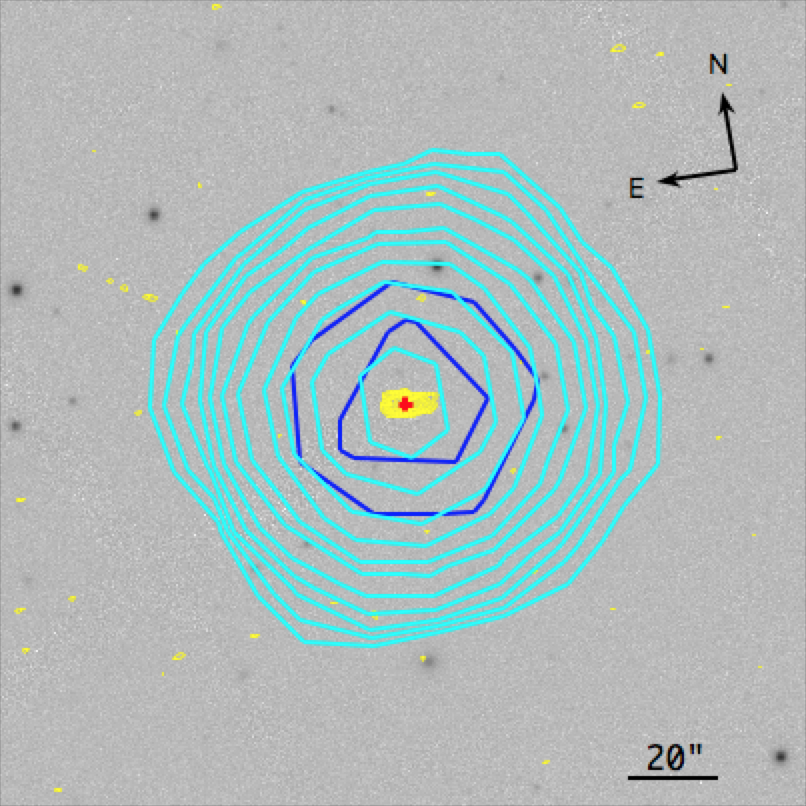}
\caption{\small 3$'$$\times$3$'$ cutout of the $i_{PS1}$ image around PSO~J191$+$86 overlaid with the radio contours from NVSS (1.4~GHz, cyan), WENSS (325~MHz, blue) and VLASS (3~GHz, yellow). In all the cases contours are spaced by $\sqrt{2}$ starting from three times the survey RMS (NVSS = 0.42 mJy beam$^{-1}$; VLASS = 160 $\mu$Jy beam$^{-1}$; WENSS = 2.7 mJy beam$^{-1}$). A zoom-in on the VLASS emission is shown in Fig. \ref{vlasszoom}.
The red cross indicates the optical position of the source.}
\label{psojradiodet}
\end{figure}
\begin{figure}[!h]
    \centering
    \includegraphics[width=8.0cm]{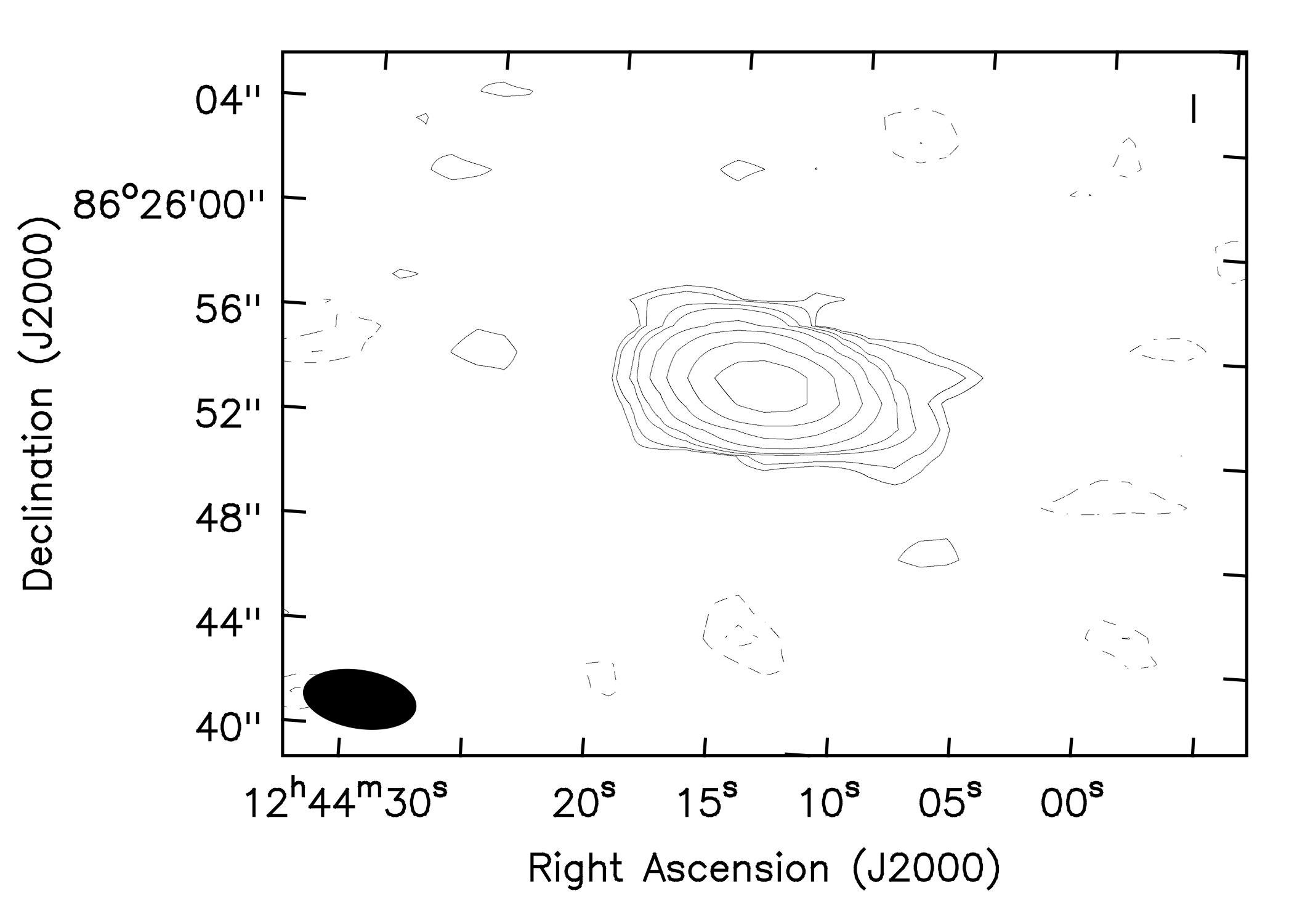}
    \caption{\small VLASS 2.1 image of PSO~J191$+$86. The contours are drawn at ($-$3, $-$2, 2, 3, 6, 9, 18, 36, 72, and 144) times the off-source RMS (160 $\mu$Jy). In the bottom left corner the beam size (4.372$''$$\times$2.241$''$) is shown. North is up, east is left.}
    \label{vlasszoom}
\end{figure}

Beside the NVSS observation at 1.4~GHz (S$_{1.4 \rm GHz}$=74.2$\pm$2.3 mJy, see also Fig. \ref{psojradiodet}), PSO~J191$+$86 has an archival detection also in the Westerbork Northern Sky Survey (WENSS, \citealt{Rengelink1997}, see Fig. \ref{psojradiodet}) at 325~MHz of 19.0$\pm$3.3~mJy.
Because of these two flux densities, \citet{Massaro2014} included PSO~J191$+$86 in the LOw frequency Radio CATalog of flat spectrum sources (LORCAT). 
Moreover the source was observed by \citet{Healey2009} at 4.85~GHz with the Effelsberg 100~m telescope. The authors measured a flux density of 33.7$\pm$1.3~mJy. 
The detections with WENSS, NVSS and Effelsberg enabled \citet{Healey2009} to classify PSO~J191$+$86 as a Gigahertz-Peaked Spectrum (GPS) source, meaning a young (<10$^{3-4}$yr) and compact (<1 kpc) radio source.
They are usually considered an early stage evolution of powerful large-scale radio galaxies of the local Universe (see \citealt{ODea2021} for a recent review). \\
PSO~J191$+$86 is also clearly detected at 3~GHz in the Very Large Array Sky Survey (VLASS, \citealt{Lacy2020}, see Fig. \ref{psojradiodet} and Fig. \ref{vlasszoom}).  
The source has been observed both during the first campaign of the first epoch (VLASS 1.1) in 2017 and the first campaign of the second epoch (VLASS 2.1) in 2020. 
Since the data of the 1.1 epoch suffers from phase errors, i.e. artifacts that alter the true size and morphology of a source, we carried out the analysis only on the 2.1 image.
We performed a single Gaussian fit, by using the task IMFIT of the Common Astronomy Software Applications package (CASA, \citealt{McMullin2007} to quantify the flux density and the size of PSO~J191$+$86 (see Fig. \ref{vlasszoom}). 
The measurements of total and peak flux densities and sizes obtained from the best fit are listed in Table~\ref{vlasspsoj1244Tab}.
The position of the source estimated by the fit is: RA = 12$^h$44$^m$13.57949$^s$ $\pm$ 0.01239$^s$, Dec = +86$^d$25$^{'}$54.04861$^{''}$ $\pm$ 0.00021$^s$, that is at 0.3$''$ from the optical PS1 position, confirming the radio-optical association of PSO~J191$+$86. 
By comparing the dimension of the source as estimated by the single Gaussian fit with the beam size (major axis = 4.372$''$, minor axis = 2.241$''$, P.A. = 80.75 deg, East of North), we concluded that the source is partially resolved.\\  
Finally PSO~J191$+$86 is not detected in the TIFR GMRT Sky Survey (TGSS, \citealt{Intema2017}) at 150 MHz. 
From the TGSS image we computed only an upper limit (at 2$\sigma$) of $\sim$4 mJy.\\
In Table~\ref{radiopsoj1244+86Tab} all the available radio flux densities of PSO~J191$+$86 are reported. 
The VLASS integrated flux density has been corrected for the 3\% of systematic uncertainty\footnote{https://science.nrao.edu/vlass/data-access/vlass-epoch-1-quick-look-users-guide}.
\begin{small}
	\begin{table*}[!h]
		\caption{\small Properties of the quick look VLASS 2.1 image of PSO~J191$+$86.}
		\label{vlasspsoj1244Tab}
		\centering
	\begin{tabular}{cccccc}
		\hline\hline
		Obs. (Rest) Freq.  &Total flux density & Peak surface brightness & major axis & minor axis & P.A.\\
		(GHz) & (mJy)  & (mJy~beam$^{-1}$) & (arcsec) & (arcsec) & (deg) \\
		(1) & (2) & (3) & (4) & (5) & (6) \\
		\hline
		3.0 (18.96) & 40.57$\pm$0.32 & 38.83$\pm$0.17 & 1.16$\pm$0.10 & <0.39  &  75$\pm$7 \\		
		\hline
	\end{tabular}
		\tablefoot{\small Col(1): observing frequency (rest-frame frequency in parenthesis); Col(2): integrated flux density; Col(3): peak surface brightness; Col(4)-Col(5): de-convolved major and minor axes of the source estimated directly from the Gaussian fit; Col(6) position angle (east of north).}
\end{table*}
\end{small}

\begin{small}
	\begin{table}[!h]
		\caption{\small Summary of the archival radio observations of PSO~J191$+$86.}
		\label{radiopsoj1244+86Tab}
		\centering
		\begin{tabular}{p{0.59cm}p{1.34cm}p{1.57cm}p{1.3cm}p{1.43cm}}
		\hline\hline
		Obs. Freq. & S$_{\nu}$ & Survey or Follow-up & Resolution & Obs. date \\
		(GHz)     & (mJy)       &   & (arcsec) &      \\
		(1) & (2) & (3) & (4) & (5) \\
		\hline
		0.150     & $<$ $\sim$4$^a$   & TGSS  & 25 & 15/03/2016 \\ 
		
		0.325     & 19.0$\pm$3.3  & WENSS & 15 &  15/06/1997  \\ 
		
		1.4       & 74.2$\pm$2.3  & NVSS & 45 & 24/06/1996 \\
		
		3.0       & 41.82$\pm$0.32 & VLASS & 2.5 & 30/08/2020 \\	
		
		4.85      & 33.7$\pm$1.3  & Effelsberg & 150 & June 2008 \\	
		\hline
		\end{tabular}
		\tablefoot{\small Col(1): Observed frequency in GHz; Col(2): integrated flux density in mJy; $a$: the TGSS value is an upper limit (at 2$\sigma$); Col(3): reference survey or dedicated follow-up; \emph{Effelsberg} refers to the observation carried out by \citet{Healey2009} with the 100m Effelsberg telescope. Col(4): angular resolution in arcsec; Col(5): Date of the observation.}
	\end{table}
\end{small}

\section{$Swift$-XRT follow-up}
\label{swift}
\begin{figure}[!h]
    \centering
    \includegraphics[width=8.0cm]{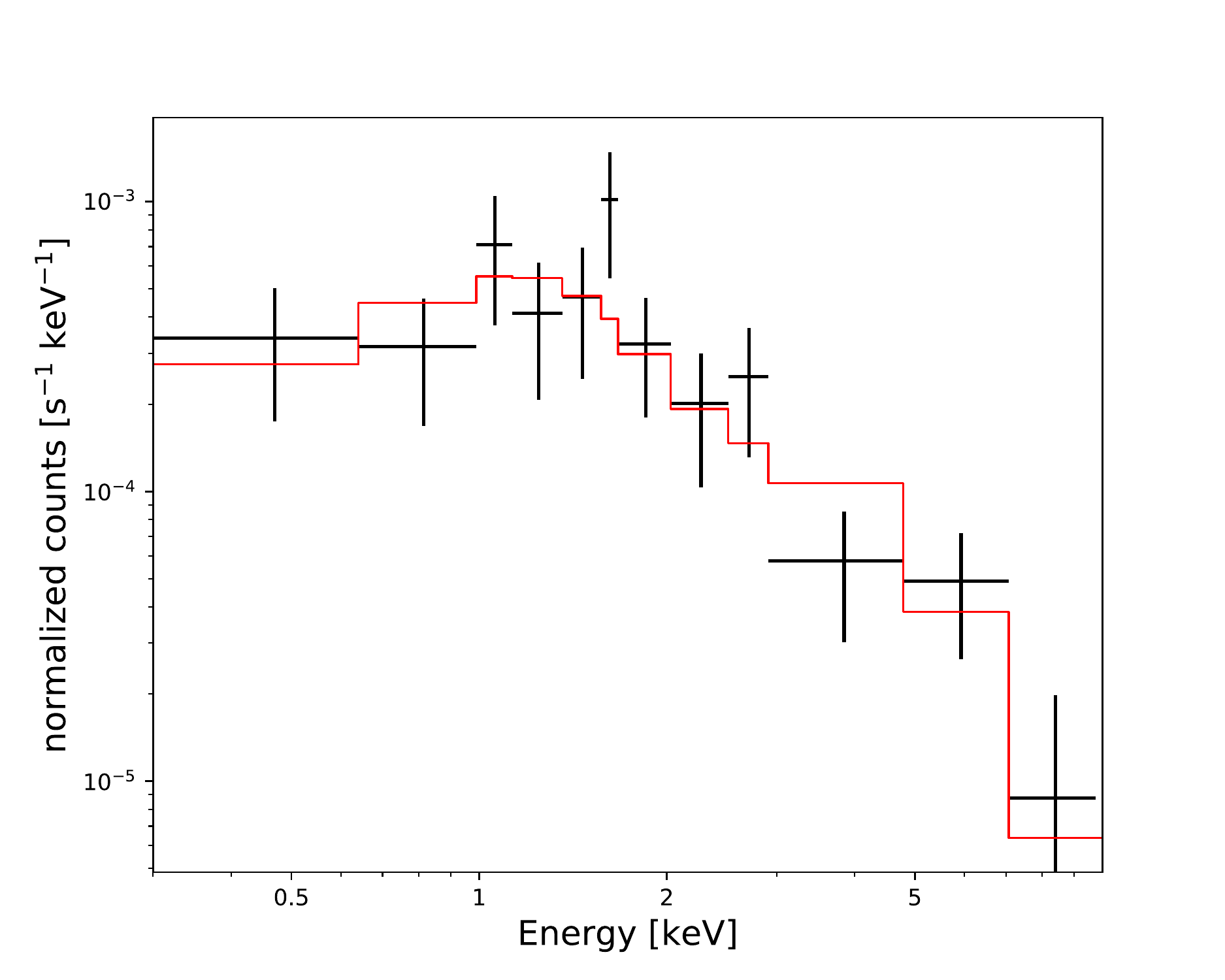}
    \caption{\small $Swift$--XRT observed spectrum of PSO~J191$+$86 modelled with a single power-law with only Galactic absorption (solid red line).}
    \label{psoj1244xrayspec}
\end{figure}

PSO~J191$+$86 was observed in the X--rays by the $Swift$--XRT telescope (target ID: 3110833; P.I. Belladitta S.). 
The observations were carried out between September, October and November 2020 and consisted of 44 segments for a total exposure time of 47.81~ks.
Data have been reduced through the standard data analysis pipeline \citep{Evans2009}, running on the UK $Swift$ Science Data Centre web page\footnote{https://www.swift.ac.uk/index.php} using HEASOFT v6.26.1.
The source is clearly detected with a total of 55 counts and with an expected background of 7 counts in [0.5-10]~keV energy band. 
The standard (PSF-fitted) $Swift$--XRT position of the source, calculated by the detected and centroid algorithm is RA = 12$^h$44$^m$12.97$^s$, Dec = $+$86$^d$25$^{'}$57$^{''}$, with an uncertainty of 3.5$''$ (90\% confidence). 
This is at 3.0$''$ from the optical PS1 position.
A standard spectral analysis has been performed using XSPEC (v.12.10.1) by fitting the observed spectrum (Fig.~\ref{psoj1244xrayspec}) with a single power law with the absorption factor fixed to the Galactic value (6.46$\times$10$^{20}$ cm$^{-2}$) as measured by the HI Galaxy map of \citet{Kalberla2005}.
We measured a photon index ($\Gamma_X$) of 1.32 $\pm$ 0.2, which implies an X--ray spectral index ($\alpha_x$ = $\Gamma_X$-1) of 0.32, and a flux of 8.11$^{+1.63}_{-1.47}$$\times$10$^{-14}$ erg~s$^{-1}$~cm$^{-2}$ in the observed [0.5-10]~keV energy band.  

\section{Analysis of Mg$\rm II$ and C$\rm IV$ broad emission lines}
\label{optirprop}
\begin{figure*}
    \centering
		{\includegraphics[width=6.0cm]{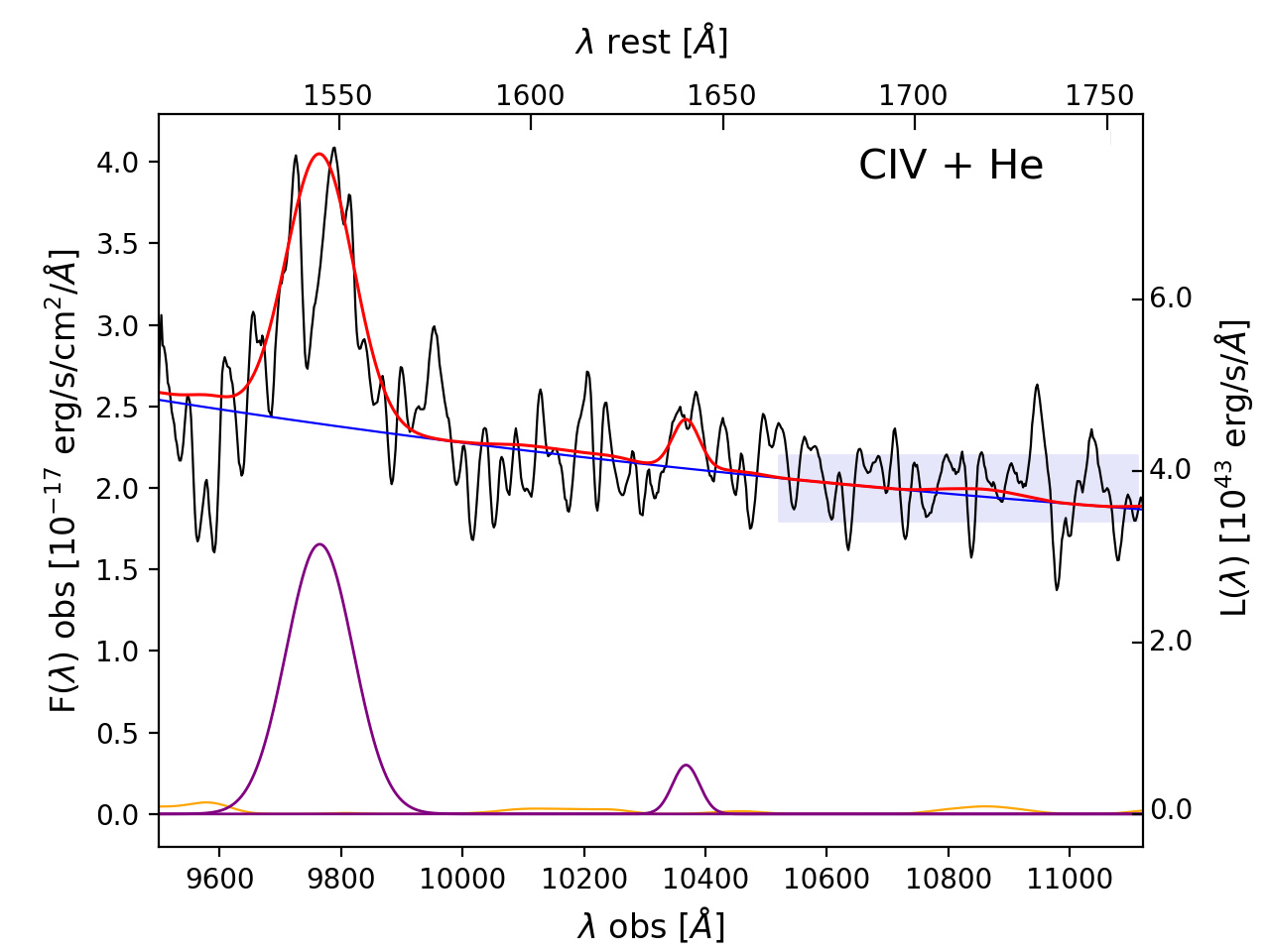}}\vspace{0.5cm}
			{\includegraphics[width=6.0cm]{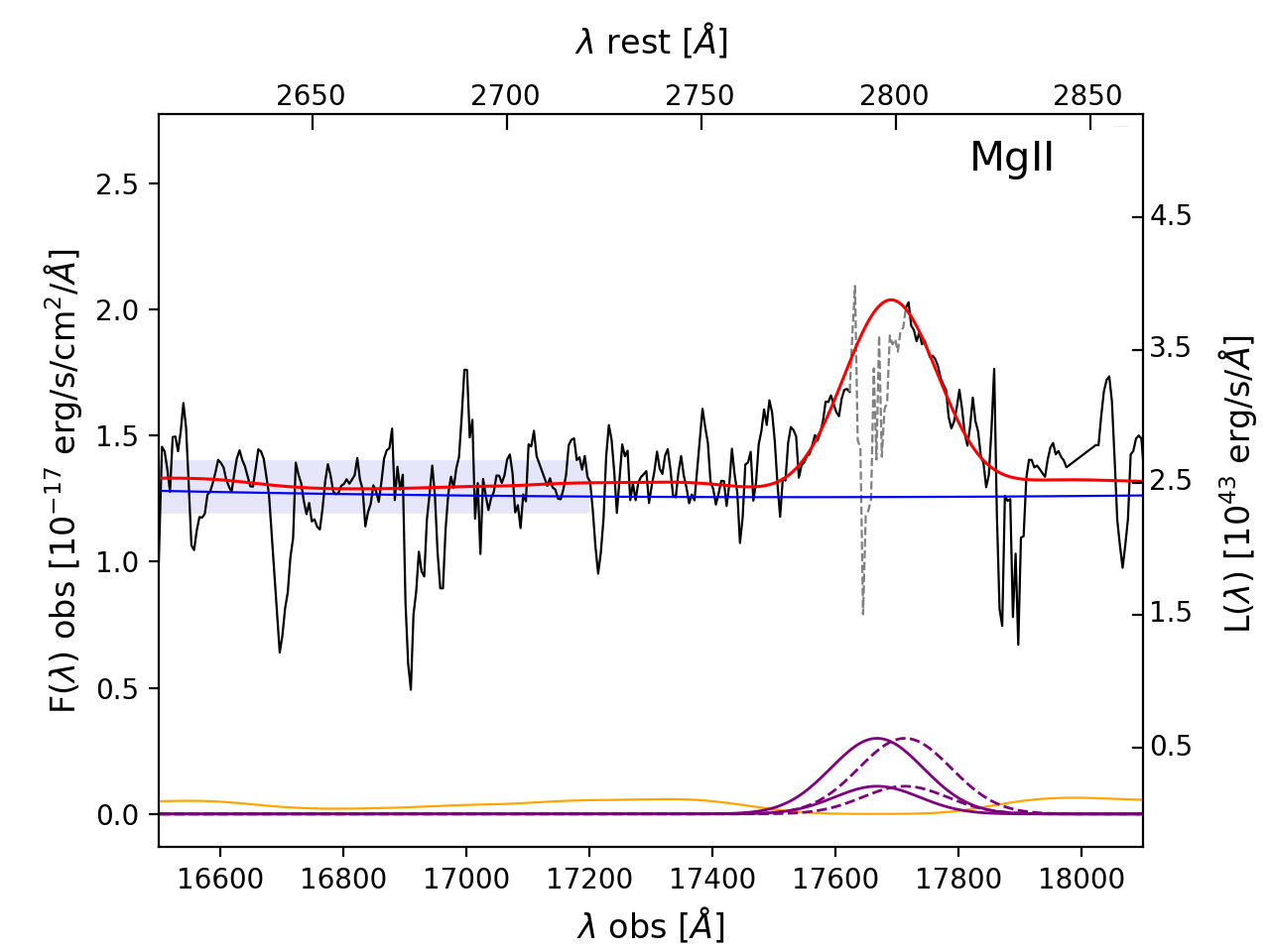}}\vspace{-0.5cm}
			{\includegraphics[width=6.0cm]{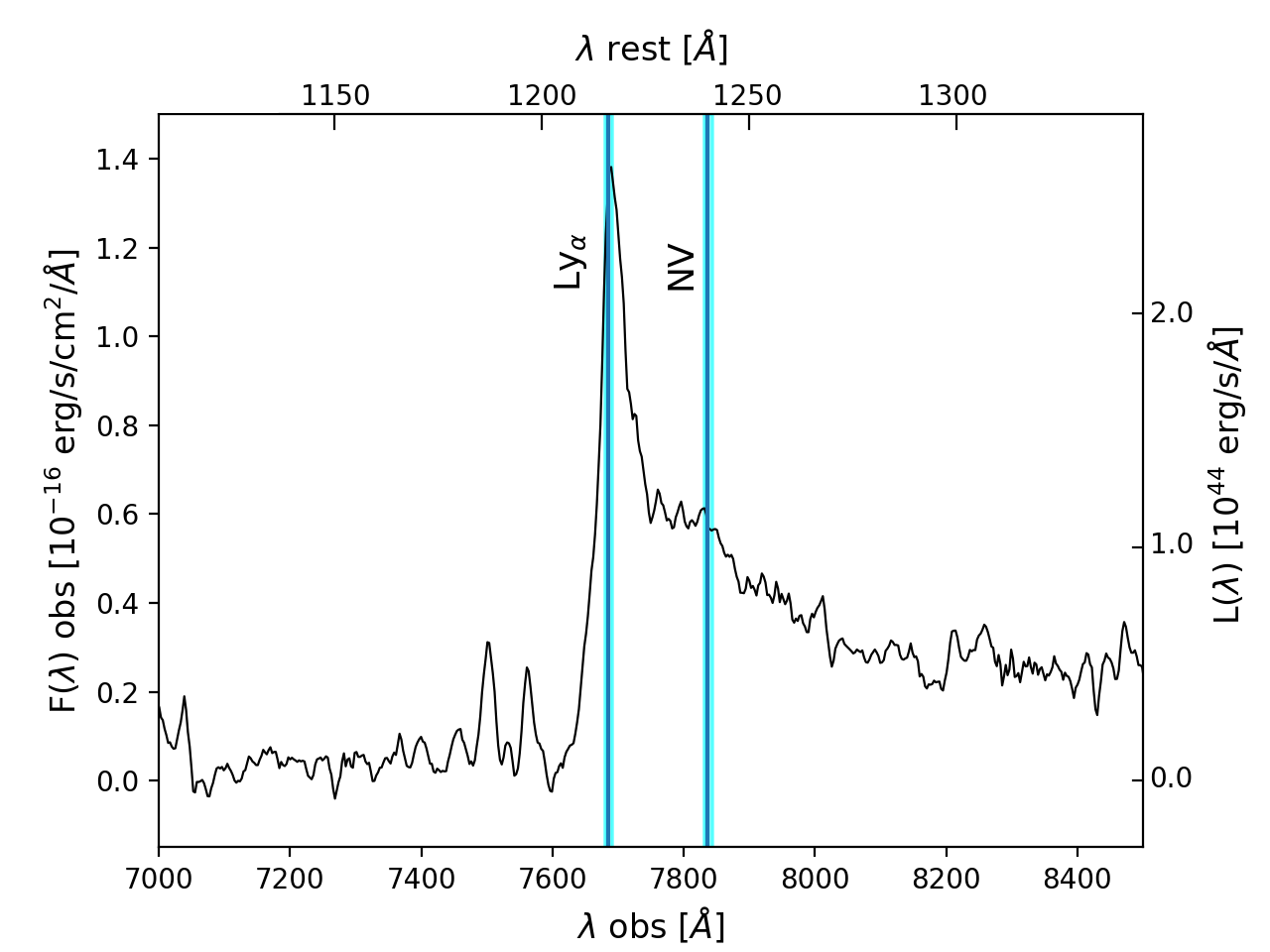}}
	\caption{\small Zoom-in on C$\rm IV$ and Mg$\rm II$ lines fit and on the Ly$_{\alpha}$+N$\rm V$ region.
	\textit{Left and central panel}: We show the total spectral fit (red line), and the different components, i.e., power law + Balmer pseudo-continuum (blue line), Fe II template (orange line, from \citealt{Vestergaard2001}, and emission lines (purple). For the Mg$\rm II$ we represent the two line of the doublet (solid Gaussians and dashed Gaussians respectively).
	Spectral regions used for the continuum fits are shown as horizontal light blue shaded areas. In the right panel, the absorption feature in the middle of the Mg$\rm II$ line is masked (grey dotted line). \textit{Right panel}: Ly$_{\alpha}$ and NV spectral region. The blue vertical lines represent the position of the two BELs based on the Mg$\rm II$ line redshift. The cyan shaded region is the uncertainty on this redshift estimate. The top x-axis reports the rest-frame wavelengths while the right axis shows the monochromatic luminosity.}
	\label{fitlines}
\end{figure*}

Among all the emission lines detected in the TNG and LBT spectra, we focused on the C$\rm IV$ and Mg$\rm II$ BELs; the latter allowed us to measure the redshift of the source and together with the C$\rm IV$ provides an estimation of the mass of the black hole hosted by PSO~J191$+$86.
To derive the properties of the two BELs (e.g., redshift, line width, line luminosity) we followed the approach of several works in the literature based on the analysis of high-z QSO spectra (e.g., \citealt{Mazzucchelli2017,Schindler2020,Banados2021,Vito2022,Farina2022}). \\ 
First, we subtracted to the spectra the continuum emission, described by a \emph{power law (f$_{pl}$ $\propto$ $\lambda$/2500$\mbox{\AA}$$^{\alpha_{\lambda}}$)}, an iron pseudo-continuum template and a Balmer pseudo-continuum.
We modeled the Fe II contribution with the empirical template of \citet{Vestergaard2001}, which is used in the derivation of the scaling relation that we later consider for estimating the black hole mass of the QSO (see Sect. \ref{bhmass}).
To be consistent with previous literature works (e.g., \citealt{Vito2022} and reference therein), with an iterative process, we convolved the iron pseudo-continuum model with a Gaussian function with the width being equal to that of the Mg$\rm II$ emission line, since it is assumed that Fe$\rm II$ emission arises from a region close to that responsible for the Mg$\rm II$ emission.
To perform the continuum fit, we chose a region of the quasar continuum free of broad emission lines and of strong spikes from residual atmospheric emission:
[10520-11110]$\mbox{\AA}$ in the LUCI1 spectrum and [15979-17200]$\mbox{\AA}$ in the LUCI2 spectrum (see Fig. \ref{fitlines}). 
We then subtracted the entire pseudo-continuum model from the observed spectra, and we modeled the two BELs with Gaussian functions. \\
BELs are usually well described by a multiple Gaussian profile (e.g., \citealt{Marziani2010,Shen2011,Shen2019,Tang2012,Karouzos2015,Rakshit2020}). 
In particular, we considered a model where each line is described by a broad plus a narrow component to account for a possible contribution from the Narrow Line Region.  
Each broad line is modelled with 1 or 2 Gaussians, while the narrow component is modelled with a Gaussian with FWHM  below 1000~km~s$^{-1}$ and larger than the spectral resolution (FWHM>$\sim$200~km~s$^{-1}$). 
In the case of Mg$\rm II$, we also considered that the line is a doublet ($\lambda$2796/2803) with an intensity ratio set equal to 1:1 (e.g., \citealt{Marziani2013}). 
For the Mg$\rm II$, we found that the narrow component is not statistically required and has no impact on the width of the broad component. 
The broad component of each line of the doublet is modelled with 2 Gaussian functions (hence 4 Gaussians in total for describing the entire line profile, see Fig. \ref{fitlines}).
Similarly, for the C$\rm IV$ we did not find a clear evidence of a narrow component. 
This is in agreement with several works in the literature, which report that the existence of a strong narrow component for C$\rm IV$ line is controversial (e.g., \citealt{Wills1993,Corbin1996,Vestergaard2002,Shen2012}).
Instead, the broad component is well fitted with a single Gaussian (Fig. \ref{fitlines}).
On the right side of the C$\rm IV$ we also fitted with a single Gaussian profile the He$\lambda$1640 emission line (see Fig. \ref{fitlines}).
The uncertainties on broad emission lines properties were evaluated through a Monte Carlo method (e.g., \citealt{Raiteri2020,Zuo2020,Diana2022}).
Each wavelength of the best fit model was randomly perturbed for 1000 times, according to a Gaussian distribution of the mean rms of the spectra computed underneath the emission lines on the pseudo-continuum subtracted spectrum. 
In this way we obtained 1000 different mock spectra of the lines profiles, from which we measured the line properties with the same procedure used on the real data.
We computed the distributions of all the line properties for these 1000 simulated spectra, and the interval that contains 68\% of the data in these distributions was taken as the statistical uncertainty on the best fit values.
From the multi-Gaussian modelling we measured the following line parameters: redshift, line width (parameterized as the FWHM), rest-frame equivalent width (REW), line flux and luminosity (see Table \ref{linesparam}). 
We adopted as the systemic redshift of PSO~J191$+$86 the one derived from the Mg$\rm II$ line, which is usually taken as the best redshift estimator for high-z objects when sub-mm and mm data are not available (e.g. \citealt{Derosa2014,Mazzucchelli2017,Schindler2020}).
We note that, despite the absorption feature in the middle, the redshift estimated by the Mg$\rm II$ line is consistent with the expected positions of the Ly$_{\alpha}$ and NV lines (see right panel of Fig.\ref{fitlines}).

\begin{small}
	\begin{table*}[!h]
		\captionsetup{width=.8\linewidth}
		\caption{\small Properties of C$\rm IV$ and Mg$\rm II$ BELs.}
		\label{linesparam}
		\centering
		\begin{tabular}{ccccccc}
			\hline\hline
			Emission line & Redshift & FWHM & REW & F$_{line}$ & L$_{line}$ & M$_{\rm BH}$ \\
			&  & km~s$^{-1}$ & $\mbox{\AA}$ & 10$^{-15}$ erg~s$^{-1}$cm$^{-2}$ & 10$^{44}$ erg~s$^{-1}$ & 10$^9$M$_{\odot}$ \\
		    (1) & (2) & (3) & (4) & (5) & (6) & (7) \\ 
			\hline \\
			C$\rm IV$ & 5.304$\pm$0.011 & 3762$_{-166}^{+124}$ & 149.75$^{+5.26}_{-4.82}$ &  2.26$^{+0.11}_{-0.08}$ & 6.49$^{+0.26}_{-0.17}$ & 2.40$_{-0.39}^{+0.42}$ \\
			Mg$\rm II$ &  5.320$\pm$0.005 & 2968$^{+339}_{-395}$  &  66.15$^{+16.11}_{-13.12}$  & 0.61$^{+0.17}_{-0.19}$  & 1.85$^{+0.54}_{-0.56}$  & 1.67$^{+0.40}_{-0.46}$  \\
			\hline
		\end{tabular}
		\tablefoot{Col (1): Emission line; Col (2): Redshift, estimated from the line peak; for the Mg$\rm II$ we report the parameters of the single doublet line. The redshift of PSO~J191$+$86 used throughout the paper is taken from the fit to the Mg$\rm II$ BEL; Col (3): FWHM in km~s$^{-1}$; Col (4): rest-frame equivalent width; Col (5) and (6): line flux and line luminosity; Col (7): SE black hole mass. Here we report the statistical uncertainty computed by propagating the error on the FWHM and continuum luminosity. In the total mass error the scaling relation scatter (0.55~dex for Mg$\rm II$ and 0.36~dex for C$\rm IV$) should be added.}
	\end{table*}
\end{small}

\subsection{Estimates of the black hole mass}
\label{bhmass}
The most common method used to compute the mass of the central BH of Type I AGN is the SE method (e.g., \citealt{Mazzucchelli2017,Shen2011,Shen2019,Diana2022}).
The underlying assumption of this technique is that the dynamics of the gas clouds is dominated by the central black hole gravitational potential. 
C$\rm IV$ and Mg$\rm II$ BELs are usually the most used lines to compute the mass of the central SMBH hosted by high-z QSOs. 
However, several works have questioned the reliability of C$\rm IV$ as a good virial mass indicator (e.g., \citealt{Sulentic2007,Trakhtenbrot2012,Bisgoni2017,Marziani2019A&A}) due to its observed blueward asymmetry and velocity shifts of the line profile with respect to low ionization lines, independently from the source orientation (e.g., \citealt{Gaskell1982,Coatman2017,Runnoe2014,Zuo2020}). These characteristics suggest that the C$\rm IV$ clouds are affected by non-gravitational effects, such as outflows, which have a significant effect on the observed emission velocity profile. 
The so-called weak emission line quasar population (WELQs, \citealt{DiamondStanic2009}, REW<10$\mbox{\AA}$) exhibits the largest blueshift ($>$3000~km~s$^{-1}$, see e.g., \citealt{Vietri2018}). 
With a REW of $\sim$150$\mbox{\AA}$, it is clear that PSO~J191$+$86 does not belong to this QSO population and, therefore, we do not expect to find a high value of blueshift for our source. 
Moreover, several works in the literature (e.g., \citealt{Marziani1996,Sulentic2007,Richards2011}) claimed that RL sources usually show lower value of blueshfit with respect to radio quiet (RQ, or \textit{non-jetted}) objects. \\
Other authors, instead, have demonstrated that there is a consistency between the SE M$_{\rm BH}$ computed from C$\rm IV$ and Balmer lines (e.g., VP06; \citealt{Greene2010,Assef2011,DallaBonta2020}).
Finally, \citet{Shen2019} report that C$\rm IV$ can still be used because on average it provides consistent black hole masses with those obtained from low ionization lines, albeit with a large intrinsic scatter. 
Therefore, we decided to compute the black hole mass also by using the C$\rm IV$, even if we considered as our best estimator the mass based on Mg$\rm II$.\\
For the Mg$\rm II$ we used the SE scaling relation presented by \citet{Vestergaard2009}, for a direct comparison with other estimates in the literature: 
\begin{equation}
\label{eqmass}
M_{\rm BH} = 10^{6.86}[\frac{\rm FWHM~(Mg \rm II)}{1000~\rm km/s}]^2 ~[\frac{\lambda L_{\lambda}~(3000 \mbox{\AA})}{10^{44}~\rm erg/s}]^{0.5}
\end{equation}
where $\lambda L_{\lambda}(3000 \mbox{\AA})$ is the monochromatic luminosity at 3000$\mbox{\AA}$ derived from the power-law model\footnote{Since the LUCI2 spectrum does not cover the wavelengths up to 3000$\mbox{\AA}$, we used the continuum near the Mg$\rm II$ line, between 2720$\mbox{\AA}$ and 2750$\mbox{\AA}$, and extrapolate it to 3000$\mbox{\AA}$ using the power-law index estimated from our data ($\alpha_{\lambda}$ = -1.51).}: $\lambda L_{\lambda}$ (3000 \mbox{\AA}) = 6.57$\pm$0.62$\times$10$^{46}$ erg s$^{-1}$. 
We obtained a black hole mass of 1.67$^{+4.18}_{-1.26}$$\times10^9$M$_{\odot}$. 
With respect to the uncertainty reported in Table \ref{linesparam}, here the error on M$_{\rm BH}$ already takes into account the intrinsic scatter of the scaling relation (0.55~dex) which is the dominant uncertainty of the black hole mass estimate.
Consistent black hole mass estimates are obtained by using different scaling relations reported in the literature, both based on the continuum luminosity (e.g., \citealt{McLure2004,Shen2012}) and on the Mg$\rm II$ line luminosity (e.g., \citealt{Shen2012}). \\
For the C$\rm IV$ we used the scaling relation of VP06, the most used in the literature: 
\begin{equation}
\label{eqmass2}
M_{\rm BH} = 10^{6.66}[\frac{\rm FWHM~(C \rm IV)}{1000~\rm km/s}]^2 [\frac{\lambda L_{\lambda}~(1350\mbox{\AA})}{10^{44}~\rm erg/s}]^{0.53}
\end{equation}
The continuum luminosity ($\lambda$L$_{\lambda}$) at 1350$\mbox{\AA}$ has been directly estimated from the TNG rest frame spectrum: 6.78$\pm$0.97$\times$10$^{46}$ erg s$^{-1}$. 
Before applying this scaling relation, we corrected the FWHM~(C$\rm IV$) value for blueshift effect, by following the prescription of \citet{Coatman2017}:
\begin{equation}
    \rm FWHM~(C \rm IV)\_{corr} = \frac{FWHM~(C \rm IV)}{\alpha \times \frac{\Delta_v}{1000 km/s} + \beta }
\end{equation}
where $\alpha$=0.41$\pm$0.02, $\beta$=0.62$\pm$0.04 and $\Delta$V is the line blueshift defined as: 
\begin{equation}
    \rm \Delta V (km~s^{-1} ) = c \times \frac{1549.48\AA - \lambda_{half}}{1549.48\AA}
\end{equation}
where $c$ is the speed of light, 1549.48$\mbox{\AA}$ is the rest frame wavelength for the C$\rm IV$ and $\lambda_{half}$ is the line centroid\footnote{The line centroid is defined as the wavelength that bisect the line in two equal part. We used the definition of \citet{DallaBonta2020}: $\lambda_{half}$ = $\frac{\int \lambda P(\lambda) d\lambda}{\int P(\lambda) d\lambda}$, where P$(\lambda)$ is the line profile.}. 
We found a $\rm \Delta V$ of 741$^{+350}_{-424}$~km~s$^{-1}$, which is small, as expected, and allowed us to infer that the virial black hole mass of PSO~J191$+$86 based on C$\rm IV$ line should not be strongly affected by blueshift effects. 
A similar value of $\rm \Delta V$ is found by using $\lambda_{peak}$ (i.e., the peak of the C$\rm IV$ line obtained by the Gaussian model in Fig. \ref{fitlines}) with respect to $\lambda_{half}$: 759$\pm$522~km~s$^{-1}$.\\
The corrected FWHM~(C $\rm IV$) is 4072$^{+180}_{-134}$~km~s$^{-1}$ and, hence, the M$_{\rm BH}$ is equal to 2.40$_{-1.40}^{+3.11}$$\times10^9$M$_{\odot}$. 
The reported error already takes into consideration the intrinsic scatter of the used scaling relation (0.36~dex; VP06).
By taking into account only the statistical uncertainty (see Table \ref{linesparam}) the black hole mass derived from C$\rm IV$ is consistent within 1$\sigma$ to the one estimated form the Mg$\rm II$ line. 

\subsection{Bolometric luminosity and Eddington ratio}
\label{lboledd}
The value of M$_{\rm BH}$ allowed us to derived the Eddington ratio of the source, which quantify how fast is the black hole accreting with respect to the Eddington limit: $\lambda_{Edd} = \frac{L_{bol}}{L_{Edd}}$, where, L$_{Edd}$ is the Eddington luminosity, the maximum luminosity beyond which radiation pressure will overcome gravity\footnote{L$_{Edd}$ = 1.26 $\frac{M_{\rm BH}}{M_{\odot}}$ $\times$ 10$^{38}$ erg s$^{-1}$} and L$_{bol}$ is the bolometric luminosity, i.e., the total energy produced by the AGN per unit of time integrated on all the wavelengths. 
To estimate it we used the continuum luminosity at 3000$\mbox{\AA}$ compute in the previous sub-section and the following bolometric correction: L$_{bol}$ = K $\times$ L$_{3000\mbox{\AA}}$, where K = 5.15 $\pm$ 1.26 \citep{Shen2008}. 
We obtained: L$_{bol}$=3.38$\pm$1.14$\times$10$^{47}$ erg s$^{-1}$.
The corresponding value of $\lambda_{Edd}$ is: = 1.55$^{+1.08~(3.50)}_{0.95~(-1.37)}$.
The uncertainty in parenthesis takes into account both the statistical error on the virial mass and the intrinsic scatter of the SE relation.\\
The SE M$_{\rm BH}$ and of $\lambda_{Edd}$ of PSO~J191$+$86 are similar to those derived for RL and RQ QSOs discovered in the same range of redshift (z=4.5-5.5; e.g., \citealt{Shen2011,Yi2014,Trakhtenbrot2021,Diana2022}). 
This similarity of masses and Eddington ratio could be likely a consequence of a selection bias, as all these high-z sources have been selected from similar optical/IR surveys.

\section{Analysis of the radio and X--ray properties}
\label{results}
\subsection{Radio spectrum}
\label{radiospec}
\begin{figure}[!h]
	\centering
	\includegraphics[width=9.5cm, height=9.5cm]{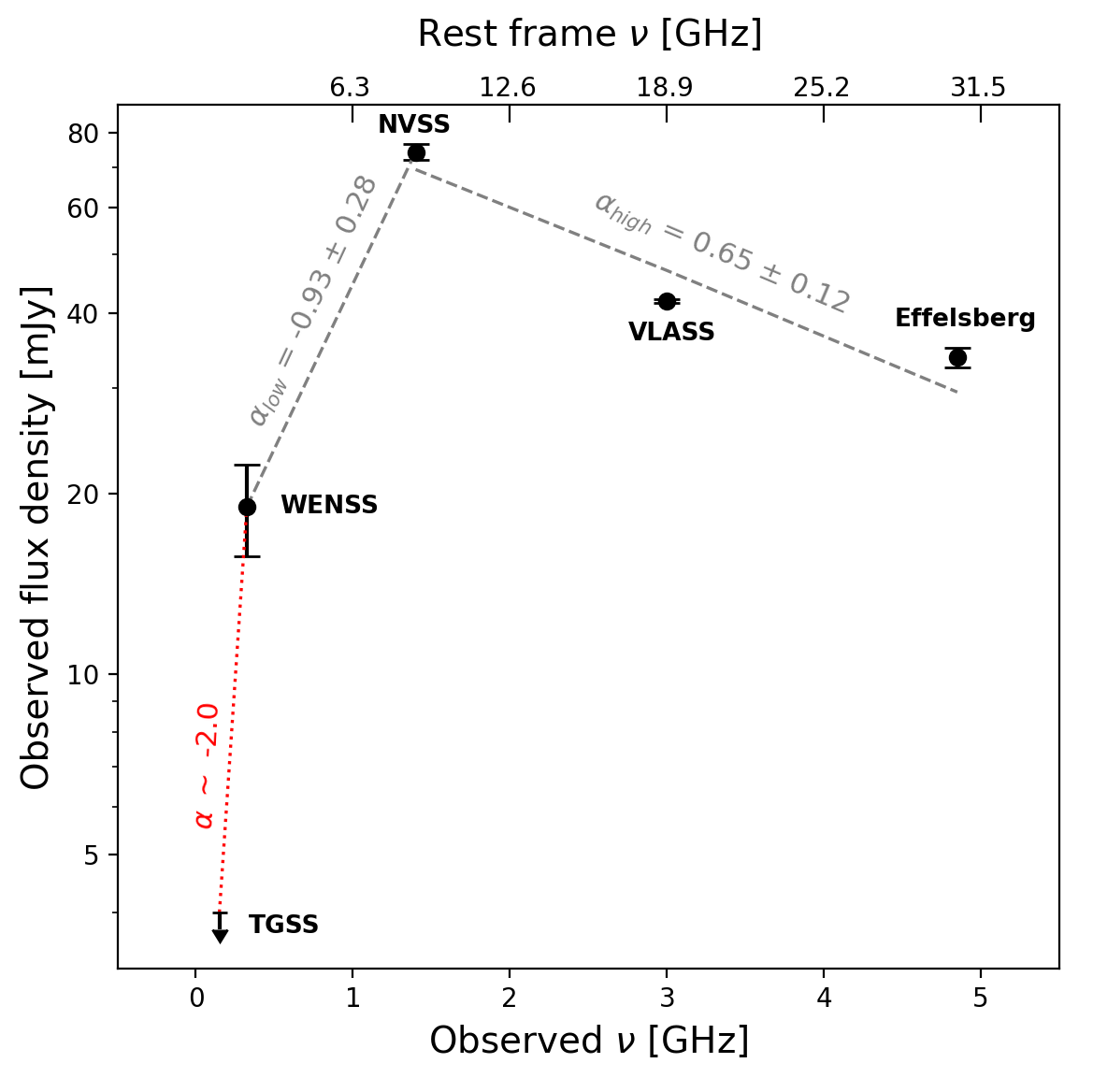}
	\caption{\small Radio flux densities as a function of the observed (bottom x-axis) and rest-frame (top x-axis) frequency of PSO~J191$+$86 from 0.150 to 4.85~GHz. The spectrum shows a peak around $\sim$1~GHz, hence the source can be classified as a GPS. 
	The corresponding indices of the low and high frequency part of the spectrum are reported together with the reference surveys. The red dashed line represents the possible spectral index according to the TGSS upper limit.}
	\label{psojrspec}
\end{figure}

From the radio flux densities listed in Table~\ref{radiopsoj1244+86Tab} we computed the radio spectral index of PSO~J191$+$86. 
We found that the radio spectrum shows a peak around $\sim$1~GHz (observed frame, Fig.~\ref{psojrspec}, $\sim$6.3~GHz in the source's rest frame). 
By assuming that the 1.4 GHz detection glimpses the turnover, we estimated the radio spectral indices below and above this potential peak and assuming a single power-law for the continuum emission (S$_\nu$ $\propto$ $\nu^{-\alpha}$).
The spectral index between 325~MHz and 1.4~GHz ($\alpha_{0.325}^{1.4}$ ) is $-$0.93$\pm$0.28 and $\alpha_{1.4}^{4.85}$ (the spectral index between 1.4 and 4.85~GHz) is 0.65$\pm$0.12.
Therefore PSO~J191$+$86 can be classified as a GPS source, according to \citet{Healey2009}.\\
Since the radio data in hand are not simultaneous (spanning over $\sim$25~years), variability may be affecting the observed spectrum, making it to appear flat (e.g., \citealt{Dallacasa2000,Orienti2020}).  
To date we have no information about possible flux density variability of the source.
A nearly simultaneous sampling of the radio spectrum (in particular below the possible turnover) is necessary.  

\subsection{Radio-loudness}
\label{RaL}
\begin{figure}[!h]
	\centering
	\includegraphics[width=9.0cm]{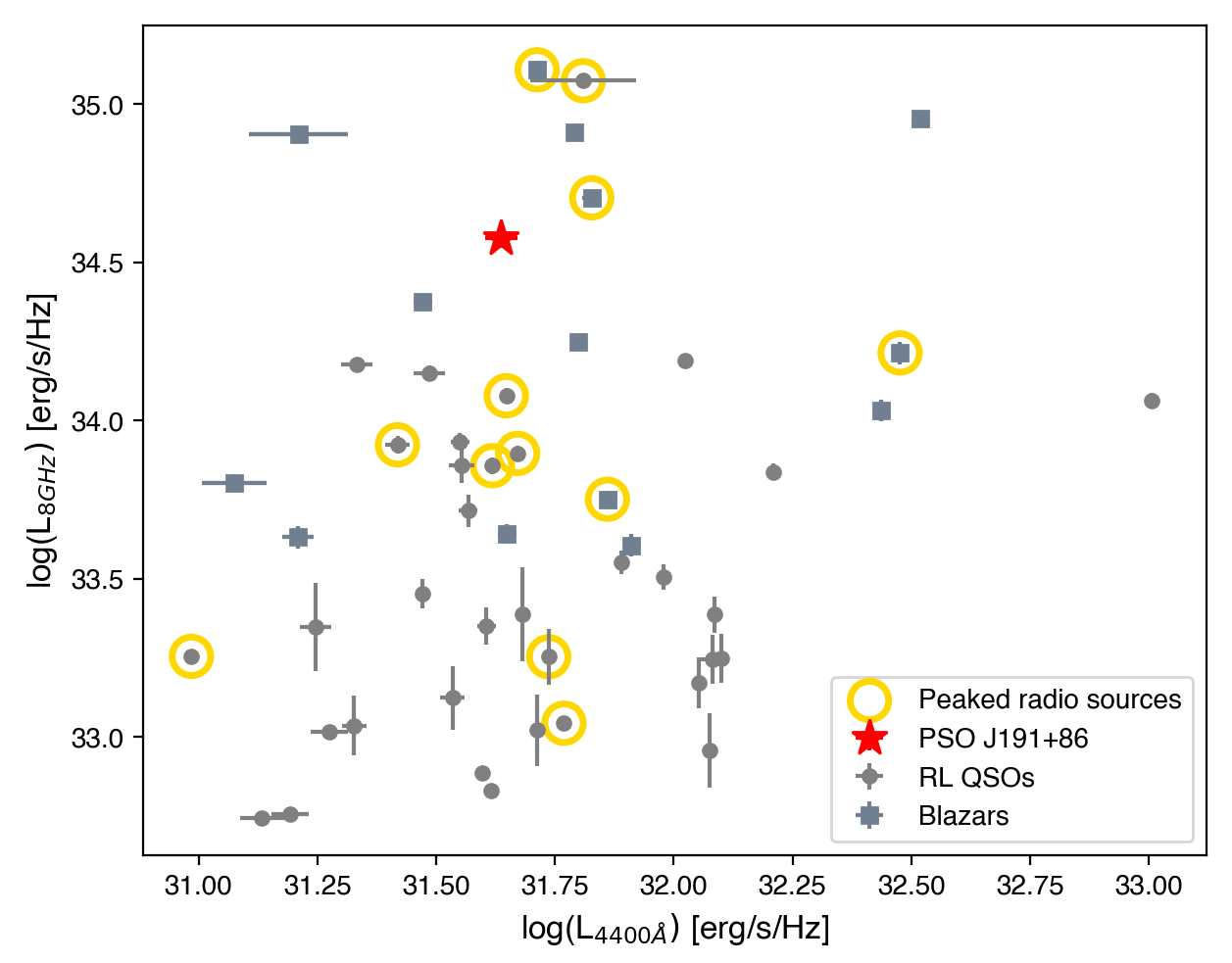}
	\caption{\small Rest-frame radio luminosity at 8 GHz vs. the rest-frame optical luminosity at 4400$\mbox{\AA}$ for PSO~J191$+$86 (red star) compared with z=4.5-5.5 radio-loud QSOs (grey points) discovered in the literature.
	Squares represent all the sources defined as blazars in the literature on the basis of their radio and/or X--ray properties.
	Yellow circles represent all the sources identified as peaked spectrum objects in the literature (\citealt{Coppejans2017,Shao2020,Shao2022}).
	PSO~J191$+$86 is one of the most luminous RL QSO at these redshifts, confirming the power of its radio emission.}
	\label{psoj1244_rcomp}
\end{figure}

From the observed radio and optical flux densities we computed the radio-loudness (R) of PSO~J191$+$86, which quantifies the level of power of the non-thermal synchrotron radio emission with respect to the thermal one originated in the accretion disk.
It is defined as the ratio between the rest-frame radio (5 GHz) and optical (blue band B at 4400$\mbox{\AA}$) fluxes \citep{Kellermann1989}: R = $\frac{S_{5 \rm GHz}}{S_{4400\AA}}$.
We estimated the flux density at 5~GHz by extrapolating the 1.4~GHz value adopting the spectral index of the optically thick part of the spectrum ($\alpha_{0.325}^{1.4}$). 
We highlight that 5~GHz in the rest-frame corresponds to 0.789 GHz in the observed frame.
The value of the flux density at 4400$\mbox{\AA}$ rest frame was derived from the LUCI2 rest frame spectrum and assuming the optical spectral index of \citet[$\alpha_{\nu}$ = 0.44]{VandenBerk2001}.
We obtained R = 337 $\pm$ 10 (Log(R) = 2.52).
This value is in agreement with that of other peaked radio sources at z$\geq$4.5, but also with the R of the high-z blazar\footnote{Blazars are RL AGN with the relativistic jets closely oriented to our line of sight (e.g., \citealt{Urry1995}).} population \citep[e.g.,][]{Belladitta2019}. 
This may suggest that the non-thermal synchrotron emission of PSO~J191$+$86 is Doppler boosted. 
Figure \ref{psoj1244_rcomp} shows the comparison between the radio luminosity at 8~GHz with the optical luminosity at 4400$\mbox{\AA}$ of PSO~J191$+$86 and of all the RL QSOs discovered in the literature at similar redshift (z=4.5-5.5)\footnote{For these RL QSOs the luminosity at 8 GHz  was computed starting from the observed flux at 1.4 GHz, and by assuming a radio spectral index of 0.75$\pm$0.25, as already done in other works \citep[e.g.,][]{Banados2015} since only few sources have a tabulated spectral index in the literature. 
However, the observed 1.4 GHz frequency corresponds to a rest-frame frequency of 7.7-9 GHz for z=4.5-5.5 sources, which is very close to 8 GHz. Therefore the lack of a measured spectral index has only a marginal impact on the computed radio luminosities.
Instead the luminosity at 4400$\mbox{\AA}$ was computed from the de-reddened z$_{PS1}$ band magnitude (the one available for all these sources) and assuming the optical spectral index of \citet[$\alpha_{\nu}$ = 0.44]{VandenBerk2001}.}. 
From this figure it is clear that PSO~J191$+$86 is one of the most luminous RL QSO at this redshift, confirming that the radio-emission is produced in a powerful jet. 
A further investigation is necessary to properly interpret PSO~J191$+$86 radio emission. 

\subsection{X--ray emission}
\begin{figure}[!h]
	\centering
	\includegraphics[width=9.0cm]{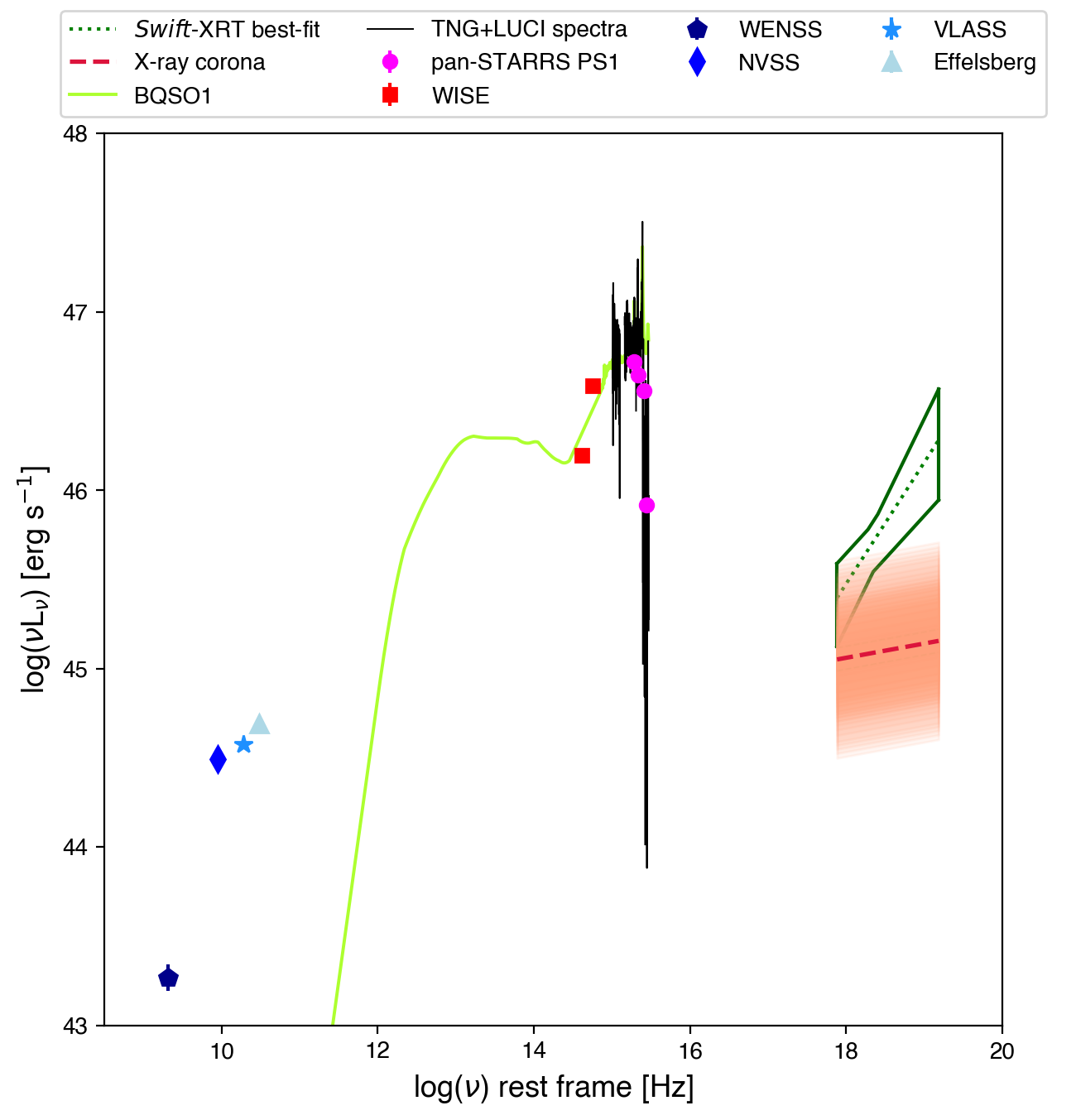}
	\caption{\small Rest-frame SED of PSO~J191$+$86 from radio to X--ray frequencies. In the X--ray we show the best fit emission in the observed [0.5-10]~keV energy band (green dotted line) with its uncertainty. We also report the TNG and LUCI spectra (in black) and a quasar template (SWIRE template library; light-grey) as guide line. The red dashed line represents the coronal emission expected from a RQ AGN with the same L$_{2500\AA}$ of PSO~J191$+$86 according to the relation of \citep{Just2007}, and the orange shaded area is the 1$\sigma$ uncertainty on this estimate. }
	\label{psoj1244sed}
\end{figure}

From the X--ray properties reported in Sect. \ref{swift} we computed an X--ray luminosity of 5.33$^{+1.54}_{-1.31}$$\times$10$^{45}$ erg~s$^{-1}$ in the [2-10]~keV energy band\footnote{L$_{02-10 keV}$ = 4$\pi$D$_{L}^2$(1+z)$^{\alpha_X -1}$F[2-10 keV], where z is the redsfhit estimated from the Mg$\rm II$ line (see Table \ref{linesparam}) and D$_{L}$ the corresponding luminosity distance computed on \url{https://www.astro.ucla.edu/~wright//CosmoCalc.html}.}.
This value clearly asseses that the X--ray emission of PSO~J191$+$86 largely overwhelms that expected from a typical hot X--ray corona and, hence, originates in a powerful jet.
This is particular evident when plotting the SED of PSO~J191$+$86 (Fig. \ref{psoj1244sed}): the X--ray emission is flatter and stronger with respect to that of RQ AGN with the same optical luminosity, according to the L$_X$-L$_{UV}$ relation of \citet{Just2007}.\\
The value of the photon index ($\Gamma_X$) computed from the $Swift-XRT$ observation (Sect. \ref{swift}) is flat (<1.5) in line with that of peaked radio sources \citep[e.g.,][]{Snios2020} and blazars \citep[e.g.,][]{Ighina2019} at high redshift. 
Therefore, we cannot exclude that the X--ray emission of PSO~J191$+$86 is dominated by a relativistic jet oriented toward the Earth, as the high value of R suggests as well. 
This is also indicated by the value of the $\tilde{\alpha}_{\rm ox}$ of PSO~J191$+$86, the ratio that quantify the relative strength of the X--ray emission with respect to the optical/UV component. 
The $\tilde{\alpha}_{\rm ox}$ is the two-point spectral index of a fictitious power law connecting 2500$\mbox{\AA}$ and 10 keV in the source rest frame \citep{Ighina2019}: $\tilde{\alpha}_{\rm ox}$ = -0.3026 $\times \log(\frac{L_{10keV}}{L_{2500\AA}})$.
The value of the monochromatic luminosity at 10~keV rest frame has been computed from the observed flux density at [0.5-10]~keV and by using the $\alpha_x$ value\footnote{L$_{10 keV}$ = $\frac{L_{0.5-10 keV} \nu_{10 keV} (-\alpha_x + 1)}{(\nu_{10 keV}^{-\alpha_x + 1} - \nu_{0.5 keV}^{-\alpha_x + 1})}$}: L$_{10~keV}$ = 2.25$^{+0.45}_{-0.41}\times$10$^{27}$ erg~s$^{-1}$~Hz$^{-1}$.
The luminosity at 2500$\mbox{\AA}$ instead has been measured from the flux density at 2500$\mbox{\AA}$ directly obtained from the LUCI 2 rest-frame spectrum: L$_{2500\AA}$ = 5.56$\pm$0.40$\times$10$^{31}$ erg~s$^{-1}$~Hz$^{-1}$.
We found an $\tilde{\alpha_{\rm ox}}$ equal to 1.329$^{+0.027}_{-0.025}$, which is similar to that of high-z blazars \citep[e.g.,][]{Ighina2019}.

\section{Discussion and conclusions}
\label{conc}
In this work we presented the discovery and multi-wavelength properties of PSO~J191$+$86, a powerful radio QSO in the early Universe (z=5.32).\\
In the radio band, PSO~J191$+$86 shows a possible peaked radio spectrum around $\sim$1 GHz in the observed frame, corresponding to $\sim$6.3 GHz in the rest frame. 
If this turnover will be confirmed, PSO~J191$+$86 will be one of the powerful (L$_{1.4 \rm GHz}$ =1.17$\times$10$^{28}$ W Hz$^{-1}$) GPS source at z$>$5 ever discovered.\\ 
By assuming that the radio spectrum is peaked, we can compute the kinetic age of PSO~J191$+$86 radio jets from the well-known correlation between peak frequency ($\nu_t$ in GHz) and projected angular size ($l$ in kpc) derived by \citet{ODea1997}: 
\begin{equation}
\label{Eqturn}
log(\nu_t) = -0.21(\pm0.05)-0.65(\pm0.05)log(l)	
\end{equation} 
Assuming that the peak of the spectrum is between 1 and 2~GHz (observed frame, i.e. $\sim$6 and $\sim$13~GHz in the rest-frame) we obtain $l$ in the range $\sim$10-30~pc (without taking into account the large scatter of the $\nu_t - l$ relation; e.g. \citealt{Nyland2020}). 
The corresponding kinetic age of the radio jets of PSO~J191$+$86 would be in the range $\sim$150-460~yr, by assuming a typical hot spot expansion velocity of 0.2c (e.g., \citealt{Giroletti2009,An&Baan2012}). 
This would make PSO~J191$+$86 one of the youngest GPS source at z$\sim$5 ever discovered. 
Therefore, a detailed study of its radio emission could shed light on the  triggering of radio activity in this distant QSO, on the formation and early stage evolution of its radio jets, on their interaction with the ambient medium and on their feedback on the host galaxy \citep[e.g.][]{Hardcastle2020}.
However the non-simultaneity of the radio data in hand does not allow us to assess the real nature of the radio spectrum, because of possible variability. 
Simultaneous observations on a wide range of frequency are necessary to confirm the possible turnover.\\
The radio loudness of the source is very high ($>$300), making PSO~J191$+$86 similar to the bulk population of blazars in the early Universe. 
Moreover the flat ($\Gamma_X$=1.32) and strong X--ray emission with respect to the optical one ($\tilde{\alpha_{\rm ox}}$ = 1.329) are similar to those of blazars.
Therefore PSO~J191$+$86 could be a GPS object with the young relativistic jets oriented closely to our line of sight. 
This is not uncommon at high-z, since the blazar Q0906$+$6930 \citep{Romani2004} at z=5.47 shows a clearly peaked radio spectrum (e.g., \citealt{Coppejans2017,Mufakharov2021}) with a turnover frequency of 6.4 GHz (observed frame), but, at the same time, Very Long Baseline Interferometry (VLBI) data found evidence of Doppler boosting \citep[e.g.][]{An2020}. These results suggest that J0906$+$6930 is a GPS source likely oriented towards the observer. Similarly a multifrequency VLBI follow-up for PSO~J191$+$86 is in progress (Belladitta et al. in prep.) to spatially resolve the pc-scale radio emission of the source and to find evidence of Doppler boosting.

\begin{acknowledgements}
We thank the anonymous referee for the useful comments and suggestions. 
	This work is based on observations made with the Large Binocular Telescope (LBT, program LBT2018AC123500-1). 
	We are grateful to the LBT staff for providing the observations for this object. 
	LBT is an international collaboration among institutions in the United States of America, Italy, and Germany.
	This work is based on observations made with the Italian Telescopio Nazionale Galileo (TNG) operated on the island of La Palma by the Fundaci\'on Galileo Galilei of the INAF (Istituto Nazionale di Astrofisica) at the Spanish Observatorio del Roque de los Muchachos of the Instituto de Astrofisica de Canarias. 
	The observations were executed by M. Pedani on a night with a short slot of DDT time available.
	This work also used data from observations with the Neil Gehrels Swift Observatory (program ID: 3110833). 
	This work made use of data supplied by the UK Swift Science Data Centre at the University of Leicester. 
	SB, AC and AM acknowledge financial contribution from the agreement ASI-INAF n. I/037/12/0 and n.2017-14-H.0 and from INAF under PRIN SKA/CTA FORECaST.
    CS acknowledges financial support from the Italian Ministry of University and Research - Project Proposal CIR01\_00010.
	This project used public archival data from the first data release of the Panoramic Survey Telescope and Rapid Response System (Pan-STARRS PS1) . 
	Pan-STARRS1 PS1 have been made possible through contributions of the institutes listed in https://panstarrs.stsci.edu.
	The NVSS data was taken by the NRAO Very Large Array. The National Radio Astronomy Observatory is a facility of the National Science Foundation operated under cooperative agreement by Associated Universities, Inc. GMRT is run by the National Centre for Radio Astrophysics of the Tata Institute of Fundamental Research. 
	WENSS is a joint project of the Netherlands Foundation for Research in Astronomy (NFRA) and Leiden Observatory. 
	VLASS data have been obtained from the Canadian Astronomy Data Centre operated by the National Research Council of Canada with the support of the Canadian Space Agency. The Canadian Initiative for Radio Astronomy Data Analysis (CIRADA) is funded by a grant from the Canada Foundation for Innovation 2017 Innovation Fund (Project 35999) and by the Provinces of Ontario, British Columbia, Alberta, Manitoba and Quebec, in collaboration with the National Research Council of Canada, the US National Radio Astronomy Observatory and Australia’s Commonwealth Scientific and Industrial Research Organisation.
	This research made use of Astropy, a community-developed core Python package for Astronomy \citep{Astropy2018}.
\end{acknowledgements}

\bibliographystyle{aa} % style aa.bst
\bibliography{ref} % your references Yourfile.bib

\end{document}